\documentclass{Interspeech}
\usepackage{multirow}
\usepackage{multicol}
\usepackage{makecell}
\usepackage{gensymb}
\usepackage{algorithm}
\usepackage{algpseudocode}
\usepackage{array}
\usepackage[]{cite}

\interspeechcameraready

\title{Thinking in Directivity: Speech Large Language Model for Multi-Talker Directional Speech Recognition\thanks{*This work was performed while author was an intern at Meta.}\thanks{$\dagger$Corresponding author.}}

\author[affiliation={1,*}]{Jiamin}{Xie}
\author[affiliation={2,\dagger}]{Ju}{Lin}
\author[affiliation={2}]{Yiteng}{Huang}
\author[affiliation={2}]{Tyler}{Vuong}
\author[affiliation={2}]{Zhaojiang}{Lin}
\author[affiliation={2}]{Zhaojun}{Yang}
\author[affiliation={2}]{Peng}{Su}
\author[affiliation={2}]{Prashant}{Rawat}
\author[affiliation={2}]{Sangeeta}{Srivastava}
\author[affiliation={2}]{Ming}{Sun}
\author[affiliation={2}]{Florian}{Metze}

\affiliation{Center for Robust Speech Systems (CRSS)}{University of Texas at Dallas}{USA}
\affiliation{}{Meta}{USA}

\email{Jiamin.Xie@UTDallas.edu, julincs@meta.com}


\usepackage{comment}

\begin{document}
\maketitle
\begin{abstract}

    Recent studies have demonstrated that prompting large language models (LLM) with audio encodings enables effective speech recognition capabilities. However, the ability of Speech LLMs to comprehend and process multi-channel audio with spatial cues remains a relatively uninvestigated area of research. In this work, we present \textit{directional-SpeechLlama}, a novel approach that leverages the microphone array of smart glasses to achieve directional speech recognition, source localization, and bystander cross-talk suppression. To enhance the model's ability to understand directivity, we propose two key techniques: serialized directional output training (S-DOT) and contrastive direction data augmentation (CDDA). Experimental results show that our proposed \textit{directional-SpeechLlama} effectively captures the relationship between textual cues and spatial audio, yielding strong performance in both speech recognition and source localization tasks.
\end{abstract}


\vspace{-2mm}
\section{Introduction}
Recent research has demonstrated that a decoder-only large language model (LLM), pre-trained on a vast text corpus, can be effectively adapted to comprehend multi-modal input, such as images and audio, by prompting the LLM with modality-specific embeddings~\cite{dubey2024llama, vicuna2023,alayrac2022flamingo,li2023blip,gong2023listen,chu2023qwen}. In particular, the integration of LLMs into speech processing has given rise to speech large language models (SLLMs), which have shown significant promise in various speech-related tasks, including automatic speech recognition (ASR), speech translation, speaker diarization, and speech synthesis~\cite{chu2023qwen,fathullah2024prompting,wu2023decoder,wang2024diarizationlm, park2024enhancing,zhang2023speechtokenizer,wang2023neural}. 


\begin{figure}[h!]
    \centering
    \includegraphics[scale=0.28]{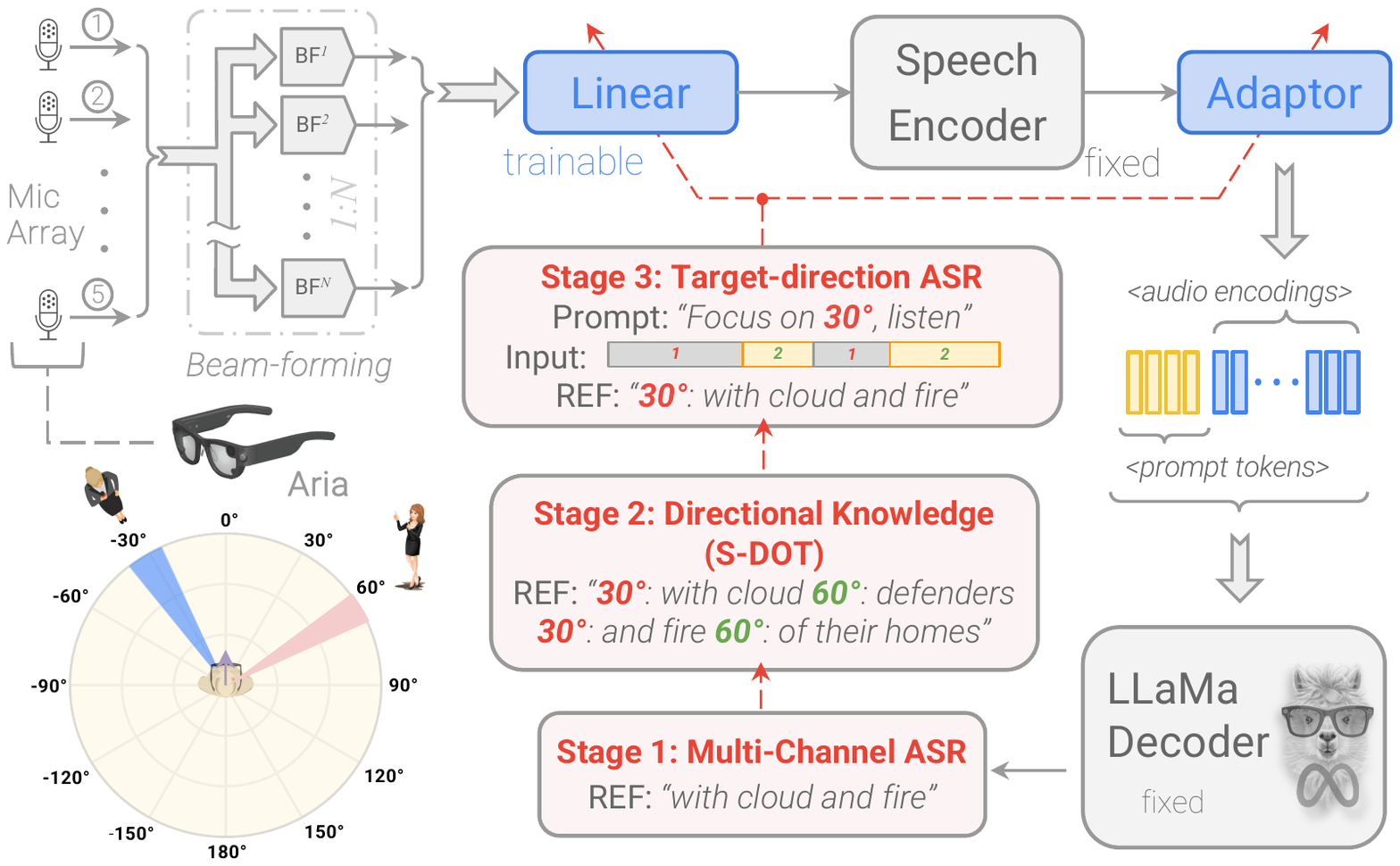}
    \caption{Illustration of Directional-SpeechLLaMA training in a multi-stage framework for multi-talker directional ASR}
      \vspace{-5mm}
    \label{summary}
  
\end{figure}

While current SLLM development has primarily focused on mono-channel data, exploring multi-channel data remains a vital research direction. 
Multi-channel audio, which records the acoustic environment from multiple spatial locations, preserves crucial spatial sound information \cite{gannot2017consolidated}. The placement of microphones determines the reception characteristics of such recordings \cite{brandstein2013microphone}. For example, the ambisonic audio captures a full 3D mapping of the sound field \cite{zotter2019ambisonics}. Microphone arrays are typically arranged in a planar configuration, enabling various beamforming techniques \cite{elko2008microphone}. Among these, the super-directive beamformer enhances sound from a specific direction \cite{elko2000superdirectional}, making it well-suited for constrained environments such as conference rooms and smart glasses \cite{xiao2016deep, ito2017probabilistic, yoshioka2018recognizing, lin2023directional,feng2023directional}. 

Moreover, multi-channel audio has been widely adopted to improve expert models designed for complex acoustic scenarios. Gu et al. \cite{gu2020enhancing} proposed learning spatial features and spectral filtering from multi-channel waveforms for overlapping speech separation. Similarly, MIMO-speech \cite{chang2019mimo} employed a multi-source neural beam-former to enhance speech for each speaker in an acoustic mixture, thereby enabling multi-talker speech separation and recognition. A recent review by Zmolikova et al. \cite{zmolikova2023neural} further emphasized the importance of multi-channel audio in target speech extraction (TSE) tasks, where it serves as both input and a cue for extracting the target speech.

Given these merits, we investigate whether a foundation model like SLLM can process multi-channel audio and, furthermore, handle multi-talker speech using spatial cues presented in text. A recent study by Zheng et al. \cite{zheng2024bat} explored spatial audio question answering, introducing a new dataset of interest and an SLLM designed to process binaural audio. However, their model was not evaluated for multi-talker speech processing. Another study by Tang et al. \cite{tang2024can} proposed an SLLM for 3D source localization, far-field ASR, and target-direction speech extraction using ambisonic audio. While their results demonstrate the potential of SLLMs for spatial audio processing, additional steps are required to extract spatial information \cite{ahonen2007teleconference}, and the target-direction ASR task is limited to distinguishing between left and right directions.

In this paper, we present a directional SLLM designed for multi-channel microphone arrays, focusing on smart glasses applications. Our approach enables the SLLM to achieve high spatial selectivity, supporting 12 discrete directions at 30-degree resolution. The proposed model leverages multiple fixed beamformers without introducing additional parameters, allowing it to perform three core tasks: multi-talker target-direction ASR, single-talker far-field ASR, and source localization.

\section{Methods}

\subsection{Multi-channel Speech Large Language Model}
Our training of multi-channel SLLM builds on recent approach that integrate speech capabilities into LLMs via audio encoders \cite{fathullah2024prompting,chu2023qwen}. The audio encoder is first trained on a large corpus of audio to acquire general acoustic representations independently of the LLM. To adapt the model to a downstream task, a specialized text prompt is constructed and concatenated with the output of the audio encoder. For ASR, a typical prompt takes the form of \textit{`Repeat after me'}, followed by the encoded audio, which is then passed through the LLM. Using paired audio-transcription data, the encoder is fine-tuned to project speech input into the continuous vector space constructed in the pre-trained LLM. Following the SpeechLLaMA approach \cite{wu2023decoder}, we developed our own model by integrating a pre-trained audio encoder with the LLaMA-3 model \cite{dubey2024llama} as the base LLM, as depicted in Figure \ref{summary}.

\subsubsection{Multi-channel beamforming front-end}
 Beamforming is a key component of our system for both direction tag prediction and cross-talk suppression. We utilize Non-Linearly Constrained Minimum Variance (NLCMV) beamforming \cite{lin2024agadir} to process the multi-channel audio inputs. These beamformers use predetermined coefficients and maintain a multi-channel structure, with each channel focusing on a distinct direction as shown in Figure \ref{summary}. The inter-channel amplitude differences in the beamformed signals effectively capture the direction-of-arrival (DOA) information, enabling our system to accurately localize and separate individual audio sources.

\subsubsection{Multi-channel (MC) SpeechLLaMA}
To process multi-channel beamformed audio input, we employ a linear projection layer and fine-tune an adapter layer of the pre-trained SpeechLLaMA model \cite{wu2023decoder}, as illustrated in Figure \ref{summary}. The audio encoder weights remain frozen to preserve its pre-trained knowledge, which operates in a high-dimensional space (1536 dimensions). The linear projection layer, inserted before the audio encoder, plays a crucial role in highlighting relevant information from different channels. It flattens the multi-channel features and projects them into the encoder's vector space, enabling the model to capture spatial cues embedded in the beamformed audio.
The adapter layer, which consists of a convolution module followed by attention layers \cite{vaswani2017attention}, was pre-trained jointly with the audio encoder for bridging general audio and the language model. We fine-tune this layer to adapt to multi-channel input, allowing the model to learn task-specific representations that reflect the spatial structure of the signal.

\subsection{Serialized Directional Output Training (S-DOT)}
The Serialized Output Training (SOT) method addresses multi-talker speech by serializing reference transcripts based on speaker start times and inserting a special speaker change token, ⟨sc⟩, between reference segments \cite{kanda2020serialized}. This token allows the model to jointly learn speaker separation and ASR, as detecting it provides implicit supervision for both tasks.
Building on this concept, we propose an extension of the SOT framework to incorporate directional information by inserting direction labels into the reference transcripts. This enables the model to jointly learn the DOA information and the speech content, effectively integrating directional awareness into the ASR process. For example, considering a set of 12 distinct directions in a plane, $D = \{\theta_{\alpha}, \theta_{\beta}, ..., \theta_{\mu} | -180\degree \leq \theta_{i} < \theta_{i+1} \leq 180 \degree, \theta_{i+1}-\theta_{i} = 30\degree\}$, the reference transcripts $R$ for the single-talker and multi-talker (a conversation between two talkers) cases are as follows: 
\begin{equation}
  R_{single} = [\langle \theta_{\alpha} \rangle, h^{\alpha}_{1}, h^{\alpha}_{2}, ..., h^{\alpha}_{T^{\alpha}}, \langle eos \rangle],
\end{equation}
\begin{equation}
  R_{multi} = [\langle \theta_{\alpha} \rangle, h^{\alpha}_{1}, ...,  h^{\alpha}_{T^{\alpha}_{i}}, \langle \theta_{\beta} \rangle, h^{\beta}_{1},  ..., h^{\beta}_{T^{\beta}_{i}}, contd],
\end{equation}
where $T^{\alpha}_{i}$ or $T^{\beta}_{i}$ denotes the transcription length of the $i$-th speech segment in direction $\alpha$ or $\beta$, respectively, and $contd$ indicates a similar labeling pattern for subsequent speech segments or the end-of-sentence token $\langle eos \rangle$. For simplicity, we assume that each speaker does not move and always speaks from a single direction during the conversation.
\subsection{Contrastive Direction Data Augmentation (CDDA)}
To model direction understanding, our objective is to achieve spatial directivity, which entails both accurate detection of the target direction and rejection of all other directions. To this end, we employ a contrastive learning \cite{khosla2020supervised} approach by augmenting the audio with a `\textit{distractor}' speaker, who speaks from an \textit{undesired} direction. This guides the model to transcribe for the desired directions while suppressing output for undesired ones, learning to distinguish directions from the entirety of the input audio. By augmenting the training examples in this manner, the model develops the ability to recognize the target direction of interest and refrain from outputting results when the input is from any non-target direction. The detailed implementation of this method is outlined in Algorithm \ref{algo1}.
\begin{algorithm}[h]
    \caption{Contrastive Direction Data Augmetation (CDDA)}
    \begin{algorithmic}[1]
        \State \textbf{Given} a direction set $D = \{\theta_{1}, \theta_{2}, ..., \theta_{12} | -180\degree \leq \theta_{i} < \theta_{i+1} \leq 180 \degree \}$, a subset $D_{target} \in D$, an utterance set $U$, and Room Impulse Response (RIR) sets $I_{d}$ for $d \in D$.
        \State \textbf{Assign} $D_{distr} = D - D_{target}$ and $U_{cdda} = [\ ]$.
        \For {$u_{1} \in U$;} 
            \State Sample $u_{2} \sim U$, $u_{3} \sim U$
            \State Sample $\theta_{1} \sim D_{target}$, $\theta_{2} \sim D_{target}$, $\theta_{3} \sim D_{distr}$
            \State Sample $r_{1} \sim I_{\theta_{1}}$, $r_{2} \sim I_{\theta_{2}}$, $r_{3} \sim I_{\theta_{3}}$
            \State $x^{j} \leftarrow $ SimulateMix$($($u_{1}, r_{1}$), ($u_{2}, r_{2}$), ($u_{3}, r_{3}$)$)$
            \State $R^{j}_{multi} \leftarrow [\langle \theta_{1} \rangle, h^{1}_{1}, ...,  h^{1}_{T^{1}_{i}}, \langle \theta_{2} \rangle, h^{2}_{1},  ..., h^{2}_{T^{2}_{i}}, contd]$
            \State Add ($x^{j}$, $R^{j}_{multi}$) $\rightarrow$ $U_{cdda}$
        \EndFor
        \State \Return $U_{cdda}$
    \end{algorithmic}
    \label{algo1}
\end{algorithm}

\begin{table*}[h!]
    \centering
    \caption{Evaluation results of direction localization and ASR, the single-talker far-field speech and direction understanding task: Word error rate (WER), direction localization accuracy (Acc.), and left ($-$) or right ($+$) localization accuracy (L/R Acc.).}
    \resizebox{2\columnwidth}{!}{
        \large
        \centering
        \renewcommand{\arraystretch}{1.2}
        \begin{tabular}{|c|c |c|*{12}{c|}c|}
        \hline
        \multirow{2}{*}{\textbf{ID}} & \multirow{2}{*}{\textbf{Model}} & \multirow{2}{*}{\textbf{Metric (\%)}} 
        & \multicolumn{13}{c|}{\textbf{Directions for Evaluation (\textdegree)}} \\ 
        \cline{4-16}
        & & & \textbf{-150} & \textbf{-120} & \textbf{-90} & \textbf{-60} & \textbf{-30} & \textbf{0} & \textbf{30} & \textbf{60} & \textbf{90} & \textbf{120} & \textbf{150} & \textbf{180} & \textbf{Avg.} \\ 
        \hline
         \multirow{2}{*}{(a)} & \multirow{2}{*}{\makecell{Whisper Large-v3 \cite{radford2023robust} \\ (w. 12 directions)}} & \multirow{2}{*}{WER} & \multirow{2}{*}{6.84} & \multirow{2}{*}{6.76} & \multirow{2}{*}{6.25} & \multirow{2}{*}{6.66} & \multirow{2}{*}{5.65} & \multirow{2}{*}{5.67} & \multirow{2}{*}{6.52} & \multirow{2}{*}{6.37} & \multirow{2}{*}{6.76} & \multirow{2}{*}{6.35} & \multirow{2}{*}{5.57} & \multirow{2}{*}{5.72} & \multirow{2}{*}{6.26} \\ 
         & & & & & & & & & & & & & & & \\
        \hline
          \multirow{2}{*}{(b)} & \multirow{2}{*}{\makecell{MC-SpeechLLaMA \\ (w. 5 directions)}} & \multirow{2}{*}{WER} & \multirow{2}{*}{5.52}	& \multirow{2}{*}{5.21} & \multirow{2}{*}{4.22} & \multirow{2}{*}{3.51} & \multirow{2}{*}{3.45} & \multirow{2}{*}{3.42} & \multirow{2}{*}{3.38}	& \multirow{2}{*}{3.52} & \multirow{2}{*}{4.59} & \multirow{2}{*}{5.25} & \multirow{2}{*}{5.37} & \multirow{2}{*}{5.48} & \multirow{2}{*}{4.41} \\
        & & & & & & & & & & & & & & & \\
         \hline
          & \multirow{3}{*}{\makecell{ (b) + S-DOT \\ (w. 5 directions)}} & WER & 5.35 & 5.21 & 4.17 & 3.60 & 3.47 & 3.52 & 3.47  & 3.66 & 4.82 & 5.20 & 6.06 & 5.63 & 4.51 \\ 
         \cline{3-16}
         (c) & & Acc. & 0 & 0 & 0 & 96 & 97 & 96 & 92 & 93 & 0 & 0 & 0 & 0 & 32 \\ 
         \cline{3-16}
         &  & L/R Acc. & 74 & 75 & 97 & 99 & 99 & 100 & 100 & 100 & 99 & 95 & 82 & 64 & 92 \\ 
         \hline
          & \multirow{3}{*}{\makecell{ (b) + S-DOT \\ (w. 12 directions)}} & WER & 3.84 & 3.77 & 3.70 & 3.72 & 3.76 & 3.84 & 3.80 & 3.76 & 3.98 & 3.91 & 4.01 & 3.93 & \textbf{3.84} \\ 
         \cline{3-16}
         (d) & & Acc. & 97 & 89 & 82 & 94 & 96 & 92 & 91 & 93 & 84 & 97 & 94 & 93 & \textbf{92} \\ 
         \cline{3-16}
         &  & L/R Acc. & 99 & 100 & 100 & 100 & 100 & 96 & 100 & 100 & 100 & 100 & 100 & 96 & \textbf{100} \\ 
         \hline
        \end{tabular}
     }
     \vspace{-5mm}
    \label{t1}
\end{table*}

\vspace{-4mm}
\subsection{Multi-stage Task Fine-tuning (Task-FT)}
We propose a three-stage cascaded framework comprising (1) Multi-channel ASR, (2) Direction-aware ASR (S-DOT), and (3) Target-direction ASR fine-tuning (Task-FT), as depicted in Figure \ref{summary}. By leveraging the hierarchical structure of these stages, we introduce a multi-stage training strategy where each stage is initialized with weights from the previous one, progressively refining the model for the final task. Starting from a pre-trained single-channel SpeechLLaMA model \cite{wu2023decoder}, our training pipeline adapts the model to multi-channel audio, enhances directional recognition, and reinforces the alignment between textual direction cues and corresponding speech. This enables the model to capture both spatial and semantic cues, improving its performance on target-direction ASR tasks.
\section{Evaluation Tasks}

\subsection{Single-talker Far-field ASR and Source Localization}
The single-talker task integrates source localization and ASR, aiming to answer the question ``\textit{what is said from where?}". The model is prompted to predict the direction of arrival in the horizontal plane and the corresponding transcript, aligning with the objectives of Stage 2 training described in Sec 2.4. The output follows the format: ``$<$\textit{direction}$>$\degree: $<$\textit{hypothesis}$>$", where the two fields represent localization and ASR predictions, respectively.
We evaluate performance using three key metrics: word error rate (WER), direction prediction accuracy (Acc.), and left/right accuracy (L/R Acc.). \textbf{WER} measures the shortest edit distance between the predicted hypothesis and the reference transcript. \textbf{Acc.} represents the proportion of correct direction predictions across all utterances. \textbf{L/R Acc.} evaluates the model's ability to generalize by checking whether the sign ($-$/$+$) of the predicted direction matches that of the ground truth, given audio from unseen directions.

\subsection{Multi-talker Target-Direction ASR}
The multi-talker task focuses on transcribing speech from a selected direction in a two-speaker conversation, addressing the question, ``\textit{From the given direction, what is said?}". The model is prompted with ``\textit{Repeat after me in $<$\textit{target}$>$\degree}" and produces output in the format ``$<$\textit{target}$>$\degree: $<$\textit{hypothesis}$>$", aligning with the objective of Stage 3 training described in Sec 2.4. We initially constrain the two speakers to occupy "different" directions and explore additional configurations in Section 3.3.

We evaluate performance using success word error rate (sWER) and success rate (SR). \textbf{sWER} measures the WER of transcriptions that correctly predict the target direction, while \textbf{SR} reflects the accuracy of the \textit{target} direction label prediction. This evaluation accounts for inevitable prediction failures: when the model fails to identify the correct direction, the hypothesis is empty, resulting in a a high WER that provide little diagnostic value, as also noted in \cite{tang2024can}. To assess generalization to unseen directions, we train the model on a restricted set of directions, specifically 30°, 60°, -30°, -60°, and 0°, and treat other directions, such as 90°, as unseen. In these cases, the model cannot predict a direction label outside the training set.
 Thus, we define success under two conditions:
\begin{itemize}
    \item \textbf{Success (audio from seen directions)}: The model's output direction label and transcription both match the target.
    \item \textbf{Success (audio from unseen directions)}: The model's output transcription matches the target transcription. 
\end{itemize}
 
Since the model's output in Stage 2 includes both direction and transcription, the target transcript can be retrieved 
either by following the exact target direction label, or, if not found, by selecting a proxy prediction based on a heuristic. To assess the model's ability to handle direction label errors while still producing accurate transcriptions, we evaluate its performance using three label error recovery schemes, as described below:
\begin{itemize}
    \item \textbf{Any Direction}: Randomly select one of the predicted direction.
    \item \textbf{Sign Match}: Select the predicted direction that has the same sign (i.e. left or right side) as the target direction.
    \item \textbf{Shortest Distance}: Select the predicted direction with the smallest cyclical distance to the target direction. For instance, in a 12-directional plane, 0\degree\ is closer than 180\degree\ to the target direction of -60\degree, with a distance of 2 vs. 4. In this case, the model's hypothesis for its predicted 0\degree\ is used as the proxy.
\end{itemize}
Note the proxy transcription may still differ from the ground-truth target transcription, resulting in evaluation failures. Additionally, we evaluate for unseen acoustic conditions, such as unseen room impulse responses (RIRs) and overlapping speech.
\vspace{-4mm}
\subsection{Extended Cases of Target-direction ASR}
To address target-direction ASR, two additional situations must be considered: (1) both speakers speak from the same direction and (2) the specified target direction in the prompt is not present in the audio. In these cases, the expected output may either include both speakers' transcriptions or be empty. However, training the model directly under such conditions may confuse it about the expected number of speakers, weakening its ability to associate direction with output content. To mitigate this, we propose appending a \textit{case label header} to the reference transcription, alongside the direction tag, enabling the model to distinguish between these complex scenarios during training.
\vspace{-2mm}
\section{Experimental Setup and Results}
We simulate 7-channel LibriSpeech (LS) datasets based on the array configuration of the Project Aria
glasses \cite{engel2023project}, using the original LS corpus \cite{panayotov2015librispeech} as the source.
Room impulse responses (RIRs) from real environments are used to model spatial diversity, generating 12 distinct directions at 30\degree\ resolution. To focus on typical conversational settings, we define five frontal directions of interest: -60\degree, -30\degree, 0\degree, 30\degree, and 60\degree.

Our model is fine-tuned from a pre-trained single-channel SLLM, following a setup and architecture similar to \cite{fathullah2024audiochatllama}. The model consists of approximately 9 billion parameters (8B from the Llama decoder + 1B from the audio encoder), with 120k fine-tuned for adaptation. We use a learning rate of 1e-4 with 4000 warm-up steps. For the single-talker task, we prepare two training configurations: a five-fold setup using the frontal direction set and a twelve-fold version covering all directions. The Whisper Large-v3 model serves as a baseline for ASR performance, using single-channel beamformed input per direction. For the multi-talker task, we prepare one-fold training data using only the frontal direction set. For evaluating unseen directions, we exclude the frontal directions and 180\degree, ensuring that each speaker is speaking from either the left or right side.

\begin{table}[t!]
  \caption{Evaluation results of target-direction ASR, the multi-talker far-field speech and direction understanding task: Success word error rate (sWER) and success rate (SR) -- direction prediction accuracy, averaging across respective directions.}
  \resizebox{\columnwidth}{!}{
    \Large
    \label{t2}
    \centering
    \renewcommand{\arraystretch}{1.1}
  \begin{tabular}{ | *{7}{c|} }
    \hline
    \centering
     \multirow{2}{*}{\textbf{ID}} & \multirow{2}{*}{\textbf{Model}} & \multirow{2}{*}{\textbf{\makecell{Label Error \\ Recovery}}} & \multicolumn{2}{c|}{\textbf{\makecell{Seen Directs. \hfill \\ Metrics (\%)}}} & \multicolumn{2}{c|}{\textbf{\makecell{Unseen Directs. \\ Metrics (\%)}}} \\
     \cline{4-7}
     & & & \makecell{\textbf{sWER}} & \makecell{\textbf{SR}} & \makecell{\textbf{sWER}} & \makecell{\textbf{SR}}\\
    \hline
    \multirow{4}{*}{(e)} & \multirow{4}{*}{\makecell{MC-SpeechLLaMA \\ + S-DOT}} & None & 4.04 & 90.4 & - & - \\
     &  & Any Direct. & 4.07 & 92.7 & 6.13 & 49.5 \\
     &  & Sign Match & 4.11 & 93.7 & \textbf{5.82} & 77.6 \\
     &  & Distance & 4.12 & 95.5 & 5.89 & \textbf{94.3} \\
    \hline
   \multirow{2}{*}{(f)} & \multirow{2}{*}{(e) + Task-FT} & \multirow{2}{*}{-} & \multirow{2}{*}{4.11} & \multirow{2}{*}{96.3} & \multirow{2}{*}{6.06} & \multirow{2}{*}{87.2} \\ 
    & & & & & & \\
        \hline
   \multirow{2}{*}{(g)} & \multirow{2}{*}{\makecell{(e) + CDDA}} & \multirow{2}{*}{\makecell{Distance}} & \multirow{2}{*}{3.94} & \multirow{2}{*}{96.2} & \multirow{2}{*}{-} & \multirow{2}{*}{-} \\ 
    & & & & & & \\
        \hline
   \multirow{2}{*}{(h)} & \multirow{2}{*}{\makecell{(e) + Task-FT \\ + CDDA}} & \multirow{2}{*}{-} & \multirow{2}{*}{\textbf{3.81}} & \multirow{2}{*}{\textbf{98.4}} & \multirow{2}{*}{-} & \multirow{2}{*}{-} \\ 
    & & & & & & \\
        \hline
  \end{tabular}
  }
  \vspace{-4mm}
\end{table}
\vspace{-2mm}
\subsection{Single-talker Task}
The evaluation results in Table \ref{t1} demonstrate the superior ASR performance of MC-SpeechLLaMA compared to the Whisper baseline, achieving an average WER of 4.41\% across all directions, despite lacking direction localization capabilities. The S-DOT method extends MC-SpeechLLaMA to incorporate spatial understanding. Comparing models (c) to (b), we observe that S-DOT achieves a 95\% direction prediction accuracy on the five seen frontal directions, with only a 0.1\% overall increase in WER, achieving source localization without degrading transcription quality. For unseen directions (e.g., 120\degree, 150\degree, 180\degree, -120\degree, -150\degree), although the model does not predict the precise labels, its outputs remain well-aligned with the ground truth, yielding a 92\% left/right accuracy. Moreover, the approach scales well to the full set of 12 directions in the training data, achieving 92\% direction prediction accuracy and a reduced WER of 3.84\% across all directions. 
\vspace{-2mm}
\subsection{Multi-talker Task}
The evaluation results in Table \ref{t2} highlight the effectiveness of applying S-DOT on the multi-talker task, achieving a raw success WER of 4.04\% with a success rate of 90.4\%. Label recovery methods further improve the success rate to 95.5\% by mitigating direction label errors, as discussed in Sec 3.2, with only a slight degradation in sWER from 4.04\% to 4.12\%. Among the recovery strategies, the shortest distance method performs best for both seen and unseen directions, suggesting the model effectively learns the notion that \textbf{\textit{sound from spatially close directions is similar}}. Fine-tuning the S-DOT model on the target task further refines its spatial directivity, increasing the success rate to 96.3\%. Comparing models (e) and (f) to (g) and (h) reveals consistent gains from applying the CDDA method, which achieves the best performance of 3.81\% success WER and 98.4\% success rate on this challenging task, albeit at the cost of neglecting unseen directions. We refer to this best-performing model, (h), as Directional-SpeechLLaMA (Base) in subsequent discussions.

\vspace{-2mm}
\subsection{Complete Cases of Multi-talker Task}
The evaluation results in Table \ref{t3} indicates that Directional-SpeechLLaMA (Base) is insufficient to learn the complete set of target-direction ASR cases using only the pre-trained LLM. However, by unfreezing one LLM decoder layer during fine-tuning, the model successfully handles all task cases, achieving an average sWER of 4.22\% and a SR of 92.3\% across the five seen frontal directions. We attribute this improvement to addressing the output mismatch between the multi-speaker target task and the single-speaker assumptions of the pre-trained model. Notably, a 0.0\% WER in the Empty output case corresponds to the model producing an empty output, which is correct when the ground-truth is also empty—such as when the prompt requests a direction (e.g., 90°) not present in the input audio (e.g., only 30° and 60° are active).
\vspace{-2mm}
\subsection{Unseen Acoustic Conditions}
The evaluation results in Table~\ref{t4} show that Directional-SpeechLLaMA remains robust under unseen RIRs from seen directions, with the success rate slightly decreasing from the topline 98.4\% to 95.0\%, while improving success WER to 3.46\%. However, the performance degrades when handling overlapping speech: at a 25\% overlap ratio, the model yields a success WER of 21.32\%, though it still maintains a 90.2\% success rate. We hypothesize this shortcoming arises from the pre-trained model not being exposed to overlapping speech, making it ineffective in recognizing content in such regions.

\begin{table}[t!]
  \caption{Evaluation results of all cases in the target-direction ASR task: Success word error rate (sWER), case label accuracy (Case Acc.), and success rate (SR) -- direction prediction accuracy, averaging across all five seen directions.}
  \resizebox{\columnwidth}{!}{
    \Large
    \label{t3}
    \centering
    \renewcommand{\arraystretch}{1.1}
  \begin{tabular}{ |*{5}{c|} }
    \hline
      \multirow{1}{*}{\textbf{Model}} & \multicolumn{1}{c|}{\textbf{\makecell{Output Cases \\ (\# of speakers)}}} & \multicolumn{1}{c|}{\textbf{\makecell{Case \\ Acc. (\%)}}} & \multicolumn{1}{c|}{\textbf{\makecell{sWER (\%)}}} & \multicolumn{1}{c|}{\textbf{\makecell{SR (\%)}}} \\
    \hline
    \multirow{2}{*}{\makecell{Directional- \\ SpeechLLaMA (Base)}} & \multirow{2}{*}{One} & \multirow{2}{*}{-} & \multirow{2}{*}{3.81} & \multirow{2}{*}{98.4} \\
    & & & & \\
   \hline
    \multirow{3}{*}{\makecell{Directional- \\ SpeechLLaMA \\ (1st decoder layer open)}} & One & 94.3 & 4.83 & 99.0 \\
    \cline{2-5}
    & Two & 91.6 & 3.61 & 91.6 \\
    \cline{2-5}
    & Empty & 86.3 & 0.00 & 86.3 \\
    \hline
  \end{tabular}
  }
  \vspace{-2mm}
\end{table}

\begin{table}[t!]
  \caption{Evaluation results of unseen acoustic condition in the target-direction ASR task: Success word error rate (sWER) and success rate (SR) -- direction prediction accuracy, averaging across all five seen directions.}
  \resizebox{0.95\columnwidth}{!}{
    \label{t4}
    \centering
    \renewcommand{\arraystretch}{1.1}
  \begin{tabular}{ |*{4}{c|} }
    \hline
      \multirow{1}{*}{\textbf{Model}} & \multicolumn{1}{c|}{\textbf{\makecell{Conditions}}} & \multicolumn{1}{c|}{\textbf{\makecell{sWER (\%)}}} & \multicolumn{1}{c|}{\textbf{\makecell{SR (\%)}}} \\
    \hline
    \multirow{3}{*}{\makecell{Directional-\\SpeechLLaMA\\(Base)}} & Seen RIR (topline) & 3.81 & 98.4 \\
    \cline{2-4}
    & Unseen RIR & 3.46 & 95.0 \\
    \cline{2-4}
    & Overlap (Ratio=25\%) & 21.32 & 90.2 \\
    \hline
  \end{tabular}
  }
    \vspace{-4mm}
\end{table}
\section{Conclusions}
This paper presents Directional-SpeechLLaMA, a speech large language model designed to process multi-channel directional audio. We propose two key techniques to infuse directional knowledge: serialized directional output training (S-DOT) and contrastive direction data augmentation (CDDA). Through a multi-stage training framework, we demonstrate effective hierarchical knowledge transfer across tasks. 
Directional-SpeechLLaMA exhibits strong performance in multi-talker speech scenarios, handling edge cases for transcribing target-direction speech based on textual spatial cues.

\bibliographystyle{IEEEtran}
\bibliography{mybib}

\end{document}